\documentclass[aps,prl,floatfix,twocolumn,reprint,amsmath,amssymb,superscriptaddress,UTF8]{revtex4-2}
\usepackage[english]{babel}

\usepackage{amsfonts}
\usepackage{mathrsfs}
\usepackage{amsmath}
\usepackage{color}
\usepackage{natbib}
\usepackage{graphicx}
\usepackage{bm}
\usepackage{amssymb}
\usepackage{xspace}
\usepackage{epstopdf}
\usepackage{dcolumn}
\usepackage{longtable}
\usepackage{array}
\usepackage{makecell}
\usepackage{tabulary}
\usepackage{multirow}
\usepackage{mhchem}
\usepackage{wrapfig}
\usepackage{pifont}  

\newcolumntype{C}[1]{>{\centering\arraybackslash}p{#1}}
\newcolumntype{L}[1]{>{\flushleft\arraybackslash}p{#1}}

\def\ie{{\it i.e.},\ }

\usepackage{multirow}
\usepackage{epsfig}
\usepackage[normalem]{ulem}
\usepackage{amsmath}
\usepackage{braket}
\usepackage{bbold}
\usepackage{natbib}

\usepackage[colorlinks=true, letterpaper=true, pdfstartview=FitV, linkcolor=blue, citecolor=blue, urlcolor=blue]{hyperref}

\makeatletter

\newcommand{\Rmnum}[1]{\expandafter\@slowromancap\romannumeral #1@}
\makeatother

\begin{document}

\title{Ridge-Spin-Layer Coupling and Emergent Ridgetronics in 2D Altermagnets}

\author{Mu Tian}
\affiliation{Key Lab of advanced optoelectronic quantum architecture and measurement (MOE), Beijing Key Lab of Nanophotonics $\&$ Ultrafine Optoelectronic Systems, and School of Physics, Beijing Institute of Technology, Beijing 100081, China}
\affiliation{International Center for Quantum Materials, Beijing Institute of Technology, Zhuhai, 519000, China}

\author{Run-Wu Zhang}
\email{zhangrunwu@bit.edu.cn}
\affiliation{Key Lab of advanced optoelectronic quantum architecture and measurement (MOE), Beijing Key Lab of Nanophotonics $\&$ Ultrafine Optoelectronic Systems, and School of Physics, Beijing Institute of Technology, Beijing 100081, China}
\affiliation{International Center for Quantum Materials, Beijing Institute of Technology, Zhuhai, 519000, China}

\author{Chaoxi Cui}
\affiliation{Key Lab of advanced optoelectronic quantum architecture and measurement (MOE), Beijing Key Lab of Nanophotonics $\&$ Ultrafine Optoelectronic Systems, and School of Physics, Beijing Institute of Technology, Beijing 100081, China}
\affiliation{International Center for Quantum Materials, Beijing Institute of Technology, Zhuhai, 519000, China}

\author{Zhi-Ming Yu}
\affiliation{Key Lab of advanced optoelectronic quantum architecture and measurement (MOE), Beijing Key Lab of Nanophotonics $\&$ Ultrafine Optoelectronic Systems, and School of Physics, Beijing Institute of Technology, Beijing 100081, China}
\affiliation{International Center for Quantum Materials, Beijing Institute of Technology, Zhuhai, 519000, China}

\author{Yugui Yao}
\affiliation{Key Lab of advanced optoelectronic quantum architecture and measurement (MOE), Beijing Key Lab of Nanophotonics $\&$ Ultrafine Optoelectronic Systems, and School of Physics, Beijing Institute of Technology, Beijing 100081, China}
\affiliation{International Center for Quantum Materials, Beijing Institute of Technology, Zhuhai, 519000, China}
\date{\today}
\begin{abstract}
Extending valleytronics from discrete points to continuous lines in momentum space transforms dispersionless bands into a controllable degree of freedom. Here we introduce ridge--spin--layer coupling (RSLC) in two-dimensional (2D) altermagnets, where a one-dimensional continuous line of dispersionless electronic states (a ridge) in momentum space locks to both spin polarization and atomic sublayer. This ridge-induced quenching of kinetic energy mimics flat-band physics, yet crucially, RSLC grants external control, allowing for layer-selective switching of ridge orientation in reciprocal space, spin-filtered transport in real space, and a distinct electric Hall response. Guided by collinear spin layer group symmetry, we identify three 2D candidate materials, namely Mg$_2$Mo$_2$(PO$_5$)$_2$, Ca(FeP)$_2$, and Mg$_2$V$_2$(SO$_5$)$_2$, each featuring a crossed-ridge structure with two ridges, one per spin channel and sublayer. Our work establishes ridgetronics as a controllable platform for direction-discriminating currents, bridging dispersionless bands with multifunctional device operation.
\end{abstract}
\maketitle

\textit{\textcolor{blue}{Introduction.}}---
The discovery of new degrees of freedom (d.o.f.) in materials has long driven progress in electronics technologies~\cite{zwanenburgSiliconQuantumElectronics2013a,zuticSpintronicsFundamentalsApplications2004c,baltzAntiferromagneticSpintronics2018c,schaibleyValleytronics2DMaterials2016b}. While early devices relied solely on electron charge~\cite{joachimElectronicsUsingHybridmolecular2000,wuGraphenesPotentialMaterial2007,wangElectronicsOptoelectronicsTwodimensional2012,fioriElectronicsBasedTwodimensional2014}, the introduction of spin d.o.f. launched spintronics and revolutionized data storage~\cite{baderSpintronics2010,wolfSpintronicsSpinBasedElectronics2001a,chappertREVIEWARTICLESInsight2007,awschalomChallengesSemiconductorSpintronics2007}. More recently, emergent d.o.f. such as the valley~\cite{gunawanValleySusceptibilityInteracting2006,takashinaValleyPolarizationSi1002006,engIntegerQuantumHall2007,xiaoValleyContrastingPhysicsGraphene2007d,rycerzValleyFilterValley2007,xuSpinPseudospinsLayered2014,suiGatetunableTopologicalValley2015,tongConceptsFerrovalleyMaterial2016d,jinImagingPureSpinvalley2018,xinValleytronicsThermoelectricMaterials2018,rycerzValleyFilterValley2007a,shkolnikovValleySplittingAlAs2002}, which originates from discrete zero-dimensional energy extrema in momentum space, have enabled valleytronics with distinctive physical responses. The search for new d.o.f. therefore remains a central theme in condensed matter physics.

\begin{figure}[h]
		\begin{center}  
			\includegraphics[width=0.5\textwidth]{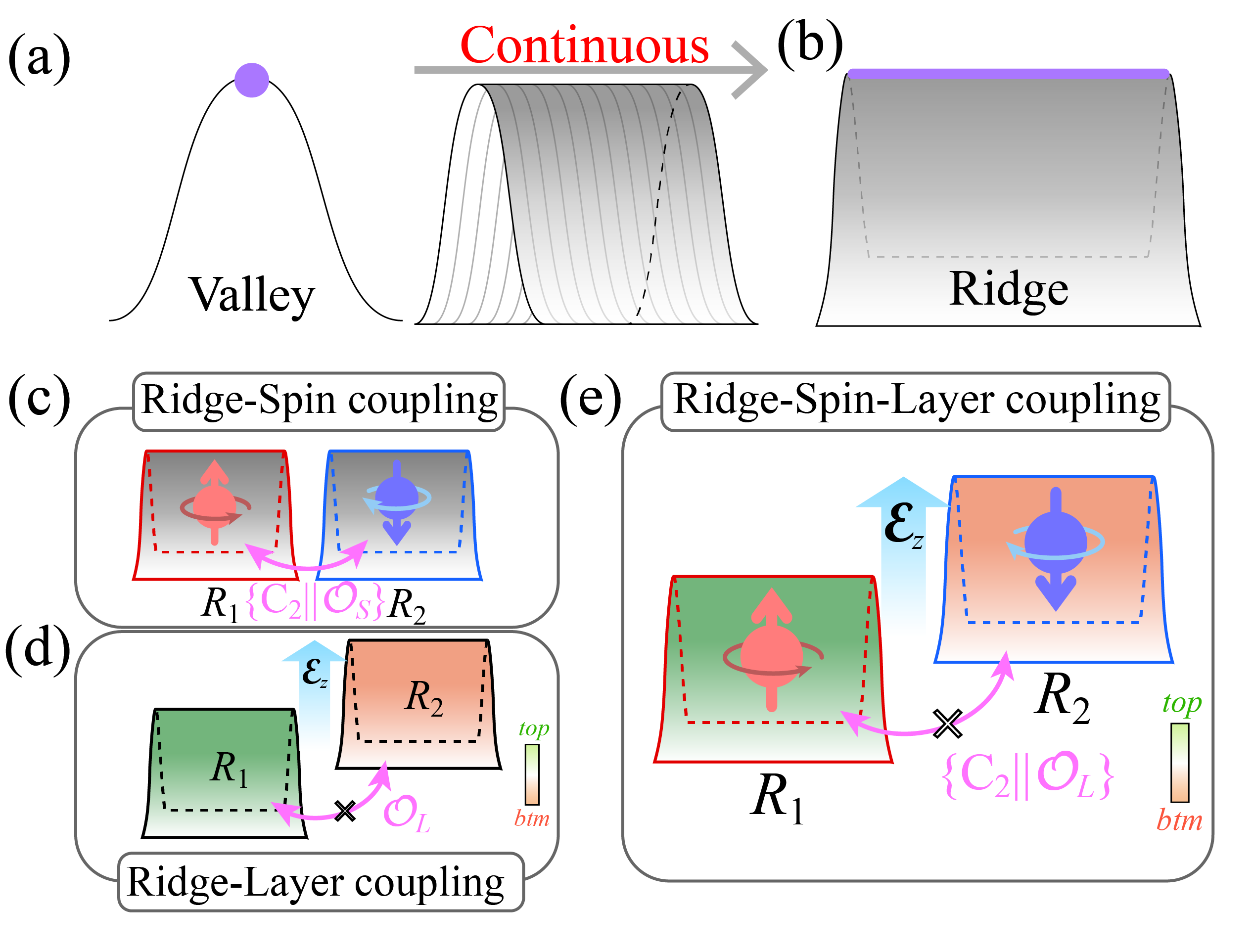}
			\caption{(a) Zero-dimensional valley (localized extremum) versus (b) ridge state (macroscopically continuous along $k_i$). 
(c) Two ridges $R_1$ and $R_2$ are connected by $\{C_2||\mathcal{O}_{S}\}$ and exhibit opposite spin polarization in the nonrelativistic case. (d) Two ridges $R_1$ and $R_2$ have opposite layer polarization, connected by $\mathcal{O}_{L}$, and tunable via an out-of-plane electric field $\mathcal{E}_z$.  (e) Combining (c) and (d): $R_1$ and $R_2$ are linked by $\{C_2||\mathcal{O}_{L}\}$, with opposite spin polarization (nonrelativistic) and opposite layer polarization.
	 } \label{fig:1}  
		\end{center}
	\end{figure}

What new physics would emerge if discrete energy extremum points (valleys) were connected into a continuous line? We term such a continuous extremum line a ``ridge''. The answer lies in the quenching of kinetic energy along the ridge direction, which intrinsically suppresses current and offers a new knob for directional transport. Unlike discrete zero-dimensional points [Fig. \ref{fig:1} (a)], certain extreme electronic states can exhibit macroscopic continuity along a specific momentum direction $k_i$, forming a ridge state [Fig. \ref{fig:1} (b)]. In such a state, the energy becomes independent of $k_i$, yielding zero group velocity $v_i$ and thus intrinsically suppressing current along that direction. This stands in contrast to conventional valleys, where electrons can propagate in all directions and current cannot be blocked along any path. As a concrete example, consider a ridge extending along $k_x$ in a two-dimensional (2D) system. Its energy takes the form:
\begin{equation}
E(\vec{k}) = f(k_y),
\label{Eq:ridge_general}
\end{equation}
which is independent of $k_x$. According to Boltzmann transport theory~\cite{pizziBoltzWannCodeEvaluation2014, Ziman_1972, scheidemantelTransportCoefficientsFirstprinciples2003}, the group velocity component $v_x = \frac{1}{\hbar}\frac{\partial E}{\partial k_x} = 0$, and the conductivity tensor is given by
\begin{equation}
\sigma_{\alpha\beta} \propto e^2 \tau v_\alpha v_\beta,
\end{equation}
where $\tau$ is the relaxation time~\cite{scheidemantelTransportCoefficientsFirstprinciples2003, madsenAutomatedSearchNew2006, book2} and $\alpha,\beta$ denote Cartesian directions. Thus, $\sigma_{xx}=0$, indicating that ridge states intrinsically suppress current along the $x$-direction, confining transport to the orthogonal direction and offering a built-in mechanism for directional control.

Beyond conceptual appeal, the practical impact of any new d.o.f. hinges on controllability, which in turn requires the ability to couple with other system knobs. Recent advances~\cite{maMultifunctionalAntiferromagneticMaterials2021b,gonzalezhernandezEfficientElectricalSpin2021,liCouplingValleyDegree2013,xiaoCoupledSpinValley2012e,grujicSpinValleyFilteringStrained2014,cuiSpinvalleyCouplingTwodimensional2021a,yuValleyLayerCouplingNew2020d,zhaoZhaoLayerValleyHallEffect2024,xuStableValleylayerCoupling2021,hanCornertronicsTwoDimensionalSecondOrder2024b,zhangQuantizedSpinHallConductivity2025d,jiangZeemanSplittingSpinvalleylayer2017,bertoniGenerationEvolutionSpin2016,jeongInterplayValleyLayer2024,zhangPredictableGateFieldControl2024b,sunDesigningSpinSymmetry2025,duanAntiferroelectricAltermagnetsAntiferroelectricity2025,zhuSlidingFerroelectricControl2025} have uncovered a rich landscape of intertwined d.o.f. in magnetic and nonmagnetic systems, including spin--valley~\cite{maMultifunctionalAntiferromagneticMaterials2021b,gonzalezhernandezEfficientElectricalSpin2021,liCouplingValleyDegree2013,xiaoCoupledSpinValley2012e,grujicSpinValleyFilteringStrained2014,cuiSpinvalleyCouplingTwodimensional2021a}, valley--layer~\cite{yuValleyLayerCouplingNew2020d,zhaoZhaoLayerValleyHallEffect2024,xuStableValleylayerCoupling2021}, spin--corner~\cite{hanCornertronicsTwoDimensionalSecondOrder2024b}, and mirror--spin~\cite{zhangQuantizedSpinHallConductivity2025d} couplings, among others~\cite{jiangZeemanSplittingSpinvalleylayer2017,bertoniGenerationEvolutionSpin2016,jeongInterplayValleyLayer2024,zhangPredictableGateFieldControl2024b,sunDesigningSpinSymmetry2025,duanAntiferroelectricAltermagnetsAntiferroelectricity2025,zhuSlidingFerroelectricControl2025}. These findings establish new paradigms for multifunctional devices. Yet a critical gap remains: how can we not only discover but also efficiently harness a new degree of freedom? This demands a general symmetry framework to identify the d.o.f. and, more importantly, to enable its active coupling to external controls, such as spin or layer, for functional ridgetronic devices.

In this Letter, we establish a complete paradigm for altermagnetic ridgetronics through a novel physical mechanism: ridge--spin--layer coupling (RSLC). Via spin layer group theory, we specify the conditions for ridge states and demonstrate their realization in altermagnets. To elucidate this mechanism, we identify three candidate materials including Mg$_2$Mo$_2$(PO$_5$)$_2$, Ca(FeP)$_2$, and Mg$_2$V$_2$(SO$_5$)$_2$ that host such ridge states. In systems with RSLC, we uncover a suite of unprecedented phenomena, including quasi-one-dimensional (Q1D) 100\% spin-polarized transport and electrically switchable spin signals. These findings establish the ridge as both a new and highly tunable electronic state, opening the field of ridgetronics as a promising frontier for controlling quantum transport in low dimensions.

\textit{\textcolor{blue}{Ridge--spin--layer coupling.}}---Taking 2D ridge altermagnets with square lattices as a representative system, we demonstrate that their physical mechanisms can lay the foundation for the generation of d.o.f. such as ridges and spins. characteristics. These d.o.f., in turn, underpin a direction-discriminating current response. This response is a hallmark of ridgetronics that extends the valleytronic paradigm into both magnetic and nonmagnetic realms.

Inspired by spintronics, where time-reversal symmetry ($\mathcal{T}$) breaking in magnetic systems enables access to the electron's spin d.o.f., we introduce a $\mathcal{T}$-broken system with only two ridges ($R_1$ and $R_2$) obtaining opposite spin polarizations.
When $R_1$ and $R_2$ are connected by the spin group symmetry $\{C_2||\mathcal{O}_{S}\}$, they coupled to up-spin and down-spin respectively [Fig.\ref{fig:1} (c)]. Among magnetic systems, altermagnets have recently emerged as the third class of magnetic materials, characterized by opposite spin sublattices connected by $\{C_2||\mathcal{O}_R\}$ ($\mathcal{O}_R$ shall only be improper or proper rotations)~\cite{smejkalConventionalFerromagnetismAntiferromagnetism2022c,smejkalEmergingResearchLandscape2022a,baiAltermagnetismExploringNew2024,fenderAltermagnetismChemicalPerspective2025,krempaskyAltermagneticLiftingKramers2024a}. Altermagnets combine properties of both ferromagnets and conventional antiferromagnets, making them an ideal platform for realizing ridge-spin coupling by naturally satisfying the required symmetry conditions.

 Within this platform, however, the two ridge states remain indistinguishable in energy and difficult to be manipulated. Introducing the layer d.o.f. overcomes this limitation, offering a versatile manipulation knob via direct coupling to out-of-plane electric field $\mathcal{E}_z$. Ridge-layer coupling is achieved when $R_1$ and $R_2$ carry opposite layer polarization $P(\boldsymbol{k})$ [see SM~\cite{SM} Eq. (7)], as permitted by $\mathcal{O}_{L}$. Applying a $\mathcal{E}_z$ breaks the symmetry $\mathcal{O}_{L2}$ and lifts the energy degeneracy between $R_1$ and $R_2$ [See Fig.~\ref{fig:1}(d)], hereby enabling precise electrical control over the direction-discriminating currents.

	
	\begin{figure}[h]
		\begin{center}  
			\includegraphics[width=0.5\textwidth]{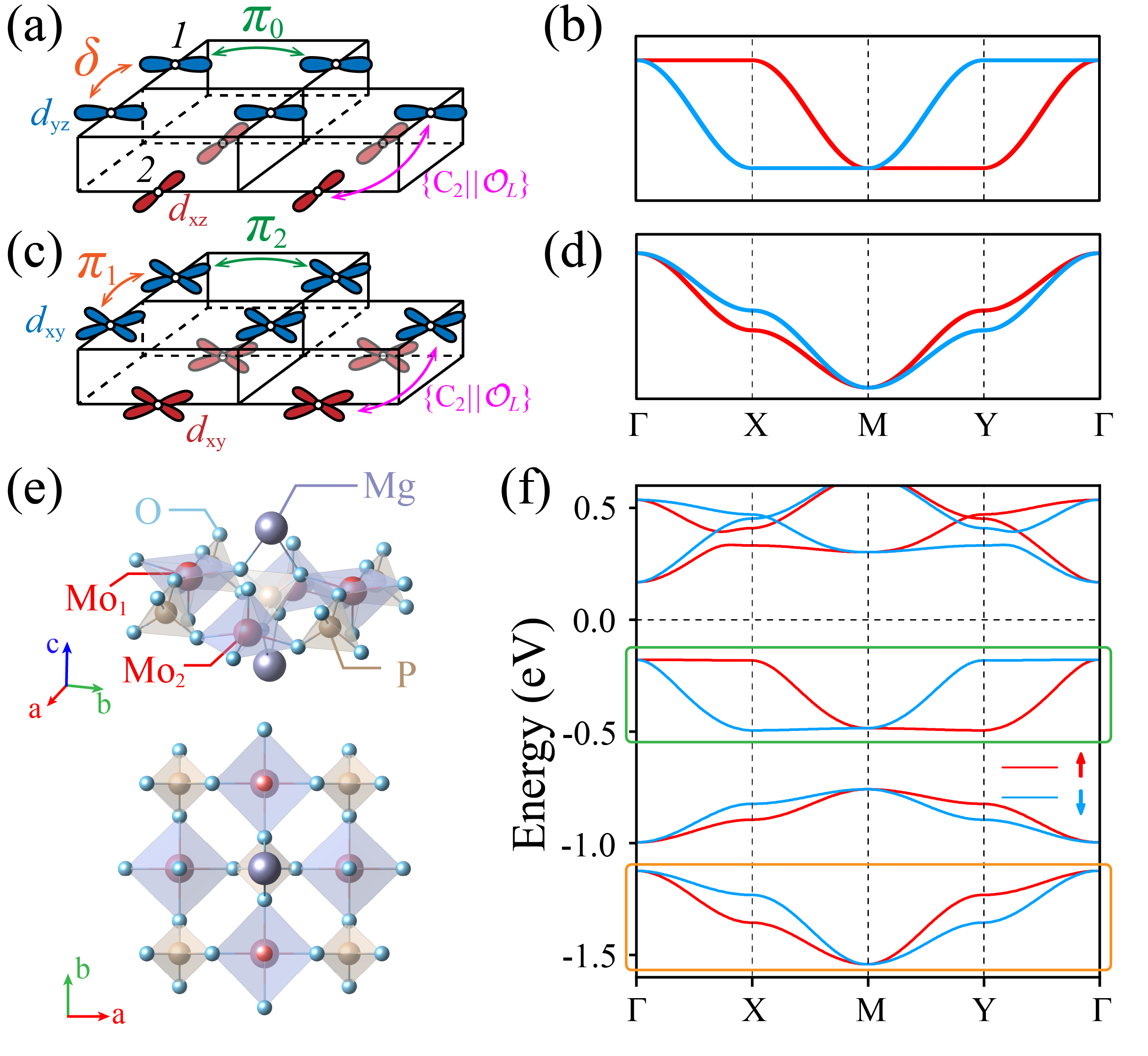}
			\caption{(a) Schematic lattice model for RSLC based on $d_{xz}$/$d_{yz}$ hopping. (b) Tight-binding band structure corresponding to the lattice in (a). (c) Schematic lattice model based on $d_{xy}$ hopping. (d) Tight-binding band structure corresponding to the lattice in (c). (e) Crystal structure of monolayer Mg$_2$Mo$_2$(PO$_5$)$_2$ in side and top views. (f) Nonrelativistic band structure of monolayer Mg$_2$Mo$_2$(PO$_5$)$_2$. Red and blue lines denote spin-up and spin-down channels, respectively. The green box highlights the $d_{xz}$/$d_{yz}$ hopping (ridge states) near the Fermi level, while the orange box marks the $d_{xy}$-based states.} \label{fig:2}  
		\end{center}
	\end{figure}

	
Altermagnets go beyond individual couplings by enabling the simultaneous integration of ridge, spin, and layer d.o.f. into a unified framework, which we term RSLC. Here, ridges $R_1$ and $R_2$ are connected by spin group symmetry $\{C_2||\mathcal{O}_{L}\}$. This novel coupling combines spin-polarized current generation from ridge--spin coupling with electrical addressability from ridge--layer coupling, introducing functionalities that transcend either alone [Fig.\ref{fig:1} (e)]. Under RSLC, each ridge locks to a specific spin-layer combination, allowing an electric field to control both ridge and spin. This multifunctional coupling thus forms the core of altermagnetic ridgetronics, offering a versatile knob for on-demand spin--layer pathways.

	\begin{table*}[t]
	\caption{List of spin layer groups (SLGs) in 2D altermagnets with square lattices that permit ridge--spin coupling, along with their corresponding spin space groups (SSGs) and the required Wyckoff positions for magnetic atoms (third column). The fourth column indicates whether each SLG additionally supports ridge--spin--layer coupling (RSLC), and candidate materials for several SLGs are listed in the final column.}.
	\begin{ruledtabular} %
		\begin{tabular}{ccccc}
			SLGs &SSGs & Wyckoff Positions&RSLC & Materials \\ 
			L.49.3.1.1 & 3.75.1.1 & 2c & \ding{55} & - \\
			L.50.3.1.1 & 3.81.1.1 & 2e,2f,2g & \ding{51} & Mg$_2$Mo$_2$(PO$_5$)$_2$,Mg$_2$V$_2$(SO$_5$)$_2$ \\
			L.51.6.1.1 & 10.83.1.1 & 2e & \ding{55} & - \\
			L.53.16.1.1 & 16.89.1.1 & 2e & \ding{55} & - \\
			L.55.23.1.1 & 25.99.1.1 & 2c & \ding{55} & - \\
			L.57.19.1.1 & 16.111.1.1 & 2e & \ding{55} & - \\ 
			L.59.23.1.1 & 25.115.1.1 & 2e,2f,2g & \ding{51} &Ca(FeP)$_2$ \\ 
			L.61.37.1.1 & 47.123.1.1 & 2f & \ding{55} & - \\
		\end{tabular}
	\end{ruledtabular}
	\label{tab:SLGs}
\end{table*}

\textit{\textcolor{blue}{Physical mechanism for RSLC.}}--Beyond proposing the novel concept of RSLC, it is equally important to establish the mechanism for its realization.  

The hopping mechanism between $t_{2g}$ ($d_{xz}$,$d_{yz}$,$d_{xy}$) orbitals provides insights for the implementation of RSLC. Specifically, the $d_{xz}$ ($d_{yz}$) orbital exhibits strong (weak) hopping along $x$ and weak (strong) hopping along $y$, respectively~\cite{chen2025orbitalhomologypt2g}. This feature serves as an important source of superconductivity in some $t_{2g}$ materials~\cite{huangExoticCooperPairing2019,kallinChiralPwaveOrder2012,raghuHiddenQuasiOneDimensionalSuperconductivity2010,linOrbitalOrderingLayered2021}. Based on the hopping characteristics, we propose the lattice model shown in Fig.~\ref{fig:2}(a), where the $d_{xz}$ and $d_{yz}$ orbitals occupy Wyckoff positions with a site multiplicity of two. The opposite magnetic moments at sites 1 and 2 result in distinct spin-channel energies for the $d_{xz}$ ($d_{yz}$) orbitals at each site, \ie $E_{i,\uparrow}^{d_{xz}}\neq E_{i,\downarrow}^{d_{xz}}$ and $E_{i,\uparrow}^{d_{yz}}\neq E_{i,\downarrow}^{d_{yz}}$, where $i=1,2$. On the other hand, since sites 1 and 2 are related by the symmetry operation $\{C_2||\mathcal{O}_{L}\}$, the following energy degeneracies hold: $E_{i,\uparrow}^{d_{xz}} = E_{j,\downarrow}^{d_{yz}},E_{i,\downarrow}^{d_{xz}} = E_{j,\uparrow}^{d_{yz}},$where $i,j = 1,2$ and $i \neq j$. For simplicity, $\{|d_{xz},\uparrow\rangle _2,|d_{yz},\downarrow\rangle _1\}$ are chosen to construct a tight-binding model as an illustration [See Fig.~\ref{fig:2}(a)]. Considering only the hopping terms shown in Fig.~\ref{fig:2}(a), the Hamiltonian can be written as
\begin{equation}
	\mathcal{H}=\varepsilon_\alpha+\left(\begin{array}{cc}
					\pi_0\mathrm{cos}k_x+\delta \mathrm{cos}k_y & 0 \\
					0 & \pi_0\mathrm{cos}k_y+\delta\mathrm{cos}k_x
				\end{array}\right).
\end{equation}
Considering $\delta \approx 0$, the band structure of this model along a suitable path is shown in Fig.~\ref{fig:2}(b). This yields the general form of the RSLC band model. 
For comparison, a similar hopping model is constructed using the basis $\{|d_{xy},\uparrow\rangle_2, |d_{xy},\downarrow\rangle_1\}$ (Fig.~\ref{fig:2}(c)). By taking the same path as in Fig.~\ref{fig:2}(b) and appropriate values of $\pi_1$ and $\pi_2$, the band structure shown in Fig.~\ref{fig:2}(d) is obtained, where the spin splitting originates from the difference between $\pi_1$ and $\pi_2$ in the system.

To guide further material searches, we establish the symmetry requirements for its material realization in the following. The above discussion establishes two fundamental prerequisites: the system must be altermagnetic in a 2D square lattice, thereby providing the necessary spin splitting and enabling orthogonal ridges; and the two ridges are connected via the symmetry operation $\{C_2||\mathcal{O}_{L}\}$.  In addition, three further criteria should be taken into account. The magnetic atoms should occupy Wyckoff positions with a site multiplicity of two, confining the unit cell to contain exactly two magnetic atoms. The magnetic moment direction must lie out of the 2D plane, preserving the symmetry of the magnetic unit cell. Furthermore, the band representation along the high-symmetry path $\Delta(0,v,0)$ in reciprocal space is 1D, ensuring that the spins of each ridge remain non-degenerate.

Applying these constraints to the 40 spin layer groups (SLGs) of altermagnetic square lattices~\cite{tianmu}, we identify 8 SLGs that satisfy all requirements for ridge--spin coupling. Within these SLGs, $\mathcal{O}_{L} \in \{C_{4z}^+, C_{4z}^-, S_{4z}^+, S_{4z}^-, C_{2[110]}, C_{2[\overline{1}10]}, M_{[110]}, M_{[\overline{1}10]}\}$. Notably, when the basis states occupy the 2e (or 2f, 2g) Wyckoff positions of SSG $P^{-1} \overline{4}^{\infty \mathrm{m}} 1$ or the 2e (or 2f, 2g) positions of SSG $P\ce{^{\text{-}1}{\text{-}4}}\ce{^{1}{m}}\ce{^{\text{-}1}{2}}\ce{^{\infty m}{1}}$, the spins may couple to opposite layers, giving rise to RSLC. In these two spin-layer groups, the orbital operator $\mathcal{O}_{L}$ belongs to the set $\{S_{4z}^+, S_{4z}^-, C_{2[110]}, C_{2[\overline{1}10]}\}$. For materials exploration, we also list the corresponding spin space groups (SSGs)~\cite{chenEnumerationRepresentationTheory2024d,jiangEnumerationSpinSpaceGroups2024,xiaoSpinSpaceGroups2024}, which are summarized together with their associated Wyckoff positions in Table~\ref{tab:SLGs}. Detailed information, including symmetry operations and Wyckoff positions is provided in Sec. II of SM~\cite{SM}.

With ridge, spin, and layer d.o.f. thus coupled, an out-of-plane electric field generates a layer-dependent electrostatic potential, enabling precise control over the ridges. This provides an ideal platform for spintronic device design.
	
	\begin{figure*}[t!]
		\centering 
		\includegraphics[width=1\textwidth]{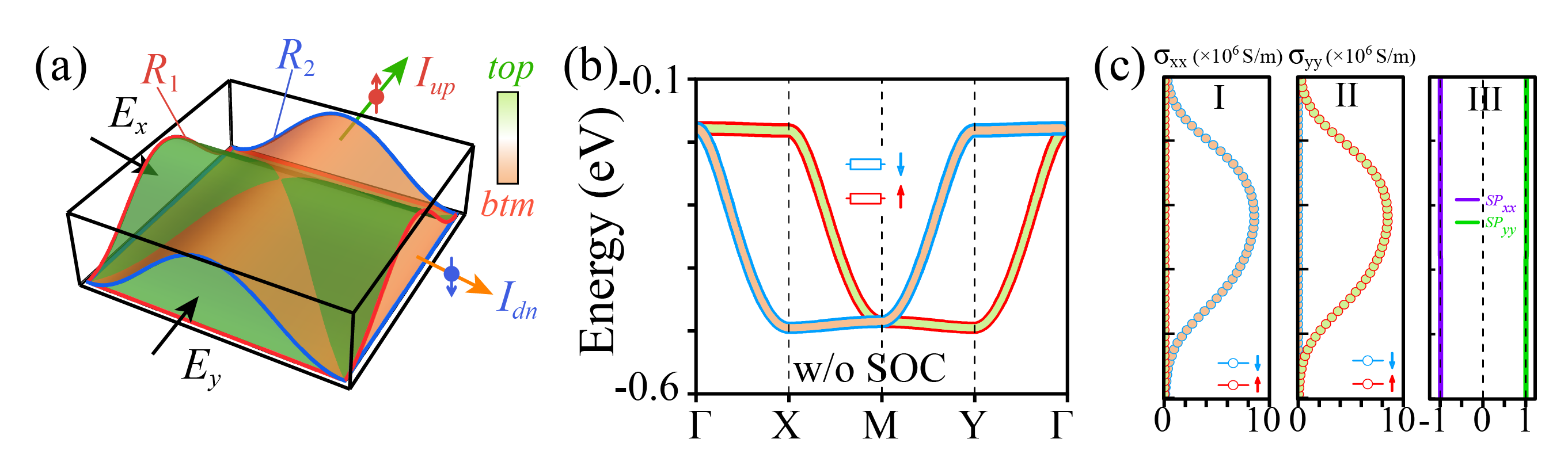}
		\caption{(a) Illustration of layer-dependent Q1D spin transport. (b) Layer-resolved analysis of the ridge states [highlighted by the green box in (b)]. Red and blue edges denote spin-up and spin-down components, respectively, while the fill color indicates layer polarization $P(\boldsymbol{k})$. The inset shows the first Brillouin zone with high-symmetry points marked. (c) From left to right: (I) calculated electric conductivity $\sigma_{xx}$ for up-spin (red-edged circles) and down-spin (blue-edged circles) channels in monolayer Mg$_2$Mo$_2$(PO$_5$)$_2$; (II) calculated electric conductivity $\sigma_{yy}$ for up-spin and down-spin channels; (III) calculated spin polarization of conductivities, $SP_{xx}$ and $SP_{yy}$. The filling color of the circles represents layer polarization $P(\boldsymbol{k})$ in all three panels.} \label{fig:3}  
	\end{figure*}

\textit{\textcolor{blue}{Candidate materials.}}---We propose several candidate materials belonging to SSG 3.81.1.1 and SSG 25.115.1.1 (see Table~\ref{tab:SLGs}), which are promising ridge materials exhibiting intrinsic RSLC. We take monolayer Mg$_2$Mo$_2$(PO$_5$)$_2$ as a representative example to demonstrate the origin of ridge states and their unique properties. The properties of additional candidates are provided in the SM~\cite{SM}.

Monolayer Mg$_2$Mo$_2$(PO$_5$)$_2$, whose parent compound is MgMo$_2$(PO$_5$)$_2$ (mp--2228679), crystallizes in a tetragonal lattice structure [space group $\overline{P}4$ (No. 81)] with optimized lattice constants $a = b = 6.56$ \AA. 
As shown in Fig. \ref{fig:2}(e), $\mathrm{Mg}^{2+}$ is bonded in a 4-coordinate geometry to four equivalent $\mathrm{O}^{2-}$ atoms. All Mg-O bond lengths are 2.03 \AA. $\mathrm{Mo}^{4+}$ is bonded to five $\mathrm{O}^{2-}$ atoms to form MoO$_5$ square pyramids (colored in blue) that share corners with four PO$_4$ tetrahedra (colored in claybank).
By comparing the energies of several possible magnetic configurations, we confirm that the ground state of the monolayer exhibits N$\acute{\mathrm{e}}$el-type magnetic order [See SM\cite{SM} for details]. 
The magnetic moments localized on the Mo sites, with a magnitude of $\sim$1.6 $\mu_B$, and align along the out-of-plane easy axis, resulting in a maximum value of the magnetic anisotropy energy of 3.23 meV/Mo~\cite{SM}.

Symmetry analysis places monolayer Mg$_2$Mo$_2$(PO$_5$)$_2$ in SSG $P^{-1}\overline{4}^{\infty m}1$. The Mo atoms occupy the Wyckoff position 2g and are therefore situated in two distinct layers, namely the top and bottom layers, as illustrated in Fig.~\ref{fig:2}(a). The top- and bottom-layer Mo atoms are related by the symmetry operation $\{C_2||S_{4z}\}$, which interchanges the spin and layer polarizations of electronic states along the $\Gamma\text{-}X\text{-}M$ and $\Gamma\text{-}Y\text{-}M$ paths. The nonrelativistic band structure of monolayer Mg$_2$Mo$_2$(PO$_5$)$_2$ is shown in Fig. \ref{fig:2}(f). The predominant basis of ridge states enclosed in the green box are $\{|d_{xz},\uparrow\rangle _{\mathrm{Mo}_2},|d_{yz},\downarrow\rangle _{\mathrm{Mo}_1}\}$, as revealed by the projected band structure of Mo $d$-orbitals in Fig. S2~\cite{SM}. This is fully consistent with our elucidation of the RSLC generation mechanism in Fig.~\ref{fig:2}(a-b). Moreover, the band structure in the orange box is dominated by the $d_{xy}$ hopping mechanism, which also corroborates the theoretical analysis in Fig.~\ref{fig:2}(c-d). These results establish monolayer Mg$_2$Mo$_2$(PO$_5$)$_2$ as an ideal ridge altermagnet hosting a RSLC.

\textit{\textcolor{blue}{Layer-dependent Q1D spin transport.}}---
2D ridge altermagnets with RSLC exhibit layer-dependent Q1D spin transport phenomenon: two orthogonal ridges are locked to opposite spins and opposite sublayers, as depicted in Fig. \ref{fig:3}(a). Near the Fermi level, conductivity is dominated by $\sigma_{xx}^{\uparrow} (\sigma_{xx}^{\downarrow})$ and $\sigma_{yy}^{\downarrow} (\sigma_{yy}^{\uparrow})$, respectively. In this setup, current $I_x$ flows along $R_1$ while $I_y$ flows along $R_2$, providing a minimal platform for ridgetronics. Of the ridge states in monolayer Mg$_2$Mo$_2$(PO$_5$)$_2$, the spin-up channel in the valence bands is predominantly localized in the bottom Mo layer, while the spin-down channel resides mainly in the top Mo layer [see Fig.~\ref{fig:3}(b)]. We calculate layer-resolved longitudinal conductivities for both spin channels in monolayer Mg$_2$Mo$_2$(PO$_5$)$_2$, are shown in Figs. \ref{fig:3}(c-I) and (c-II). 
	
The system exhibits strictly spin-anisotropic conductivity: only $\sigma_{xx}^{\downarrow}$ and $\sigma_{yy}^{\uparrow}$ are nonzero, in excellent agreement with theory. We define the spin polarization of conductivities as:
	\begin{equation}
		\mathrm{SP}_{\hat{n} \hat{n}}=\frac{\sigma_{\hat{n} \hat{n}}^{\uparrow}-\sigma_{\hat{n} \hat{n}}^{\downarrow}}{\sigma_{\hat{n} \hat{n}}^{\uparrow}+\sigma_{\hat{n} \hat{n}}^{\downarrow}}.
	\end{equation}
As shown in Fig. \ref{fig:3} (c-III), $SP_{xx}$ and $SP_{yy}$ reache -1 and 1, respectively, within the altermagnetic ridge state energy window. Thus, the ridge state generates fully spin-polarized currents with opposite spin orientations along orthogonal directions.
	
RSLC also gives rise to intrinsically layer-dependent Q1D spin transport. Symmetry analysis shows that monolayer Mg$_2$Mo$_2$(PO$_5$)$_2$ exhibits RSLC via two ridges along the $\Gamma\text{-}X\text{-}M$ and $\Gamma\text{-}Y\text{-}M$ paths, interconnected by $\{C_2 || S_{4z}\}$, which swaps both spin and layer polarizations. This symmetry-enforced coupling locks spin and layer d.o.f. to orthogonal ridge directions, underpinning the observed layer-dependent Q1D spin transport.

\begin{figure*}[t!]
	\centering 
	\includegraphics[width=1\textwidth]{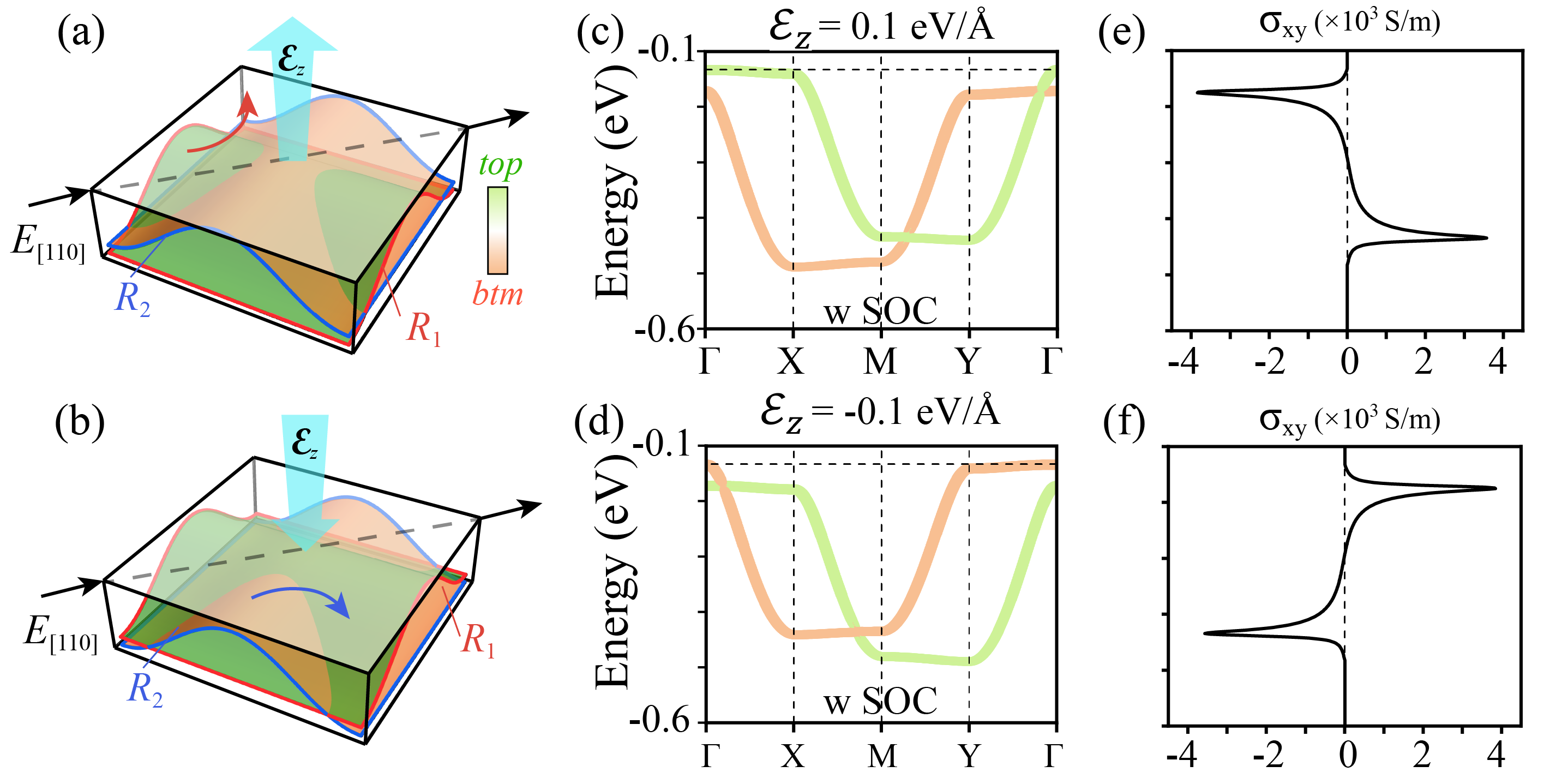}
	\caption{(a,b) Illustration of the layer-dependent electric Hall effect under opposite electric fields $\mathcal{E}_z $. (c,d) Relativistic ridge state of monolayer Mg$_2$Mo$_2$(PO$_5$)$_2$ under applied electric fields $\mathcal{E}_z=$ 0.1 and -0.1 eV/\AA~applied separately. (e,f) Anomalous hall conductivity of monolayer Mg$_2$Mo$_2$(PO$_5$)$_2$ under the same field conditions as in (c,d).} \label{fig:4}  
\end{figure*}

\textit{\textcolor{blue}{Layer-dependent electric Hall effect.}}---
The electric Hall effect (EHE)~\cite{cuiElectricHallEffect2025} in monolayer Mg$_2$Mo$_2$(PO$_5$)$_2$ arises from an electric gate field rather than a magnetic field. RSLC allows the electric field to selectively manipulate layer-polarized ridges, producing a layer-dependent Hall current without an external magnetic field. This mechanism offers an efficient route for controlling 2D ridge altermagnets via electric gate tunability.
		
In our system, two ridges are connected via $\{C_2||S_{4z}\}$ in the nonrelativistic case, with out-of-plane magnetic moments as ground state (MAE = 6.6~$\mathrm{meV}$). The monolayer Mg$_2$Mo$_2$(PO$_5$)$_2$ thus possesses $S_{4z}\cal{T}$ symmetry in the relativistic case, forbidding the intrinsic Hall conductivity $\sigma_{xy}$ but allowing the intrinsic EHE coefficient $\chi_{xy}$. Owing to SLC, the ridge states are tunable by the gate field ${\cal{E}}_z$, as illustrated in Figs.~\ref{fig:4}(a-b).

First-principles calculations agree excellently with symmetry analysis. Without ${\cal{E}}_z$, the ridges are degenerate and $\sigma_{xy}=0$. Applying ${\cal{E}}_z = 0.1 \text{eV}/$\AA~lifts the degeneracy between $R_1$ and $R_2$[Fig.~\ref{fig:4}(c)], and induces a sizable $\sigma_{xy}$ [Fig.~\ref{fig:4}(e)]. No interfering bands appear between -0.5 eV to -0.1 eV, yielding a pronounced EHE peak. Reversing ${\cal{E}}_z$ reverses the sign of $\sigma_{xy}$ [Fig.~\ref{fig:4}(f)], enabling control over the Hall current direction.

\textit{\textcolor{blue}{Conclusion.}}---
In summary, we have introduced the concept of ridge altermagnets as 1D dispersionless electronic continua in momentum space and realized it in 2D altermagnets via a novel RSLC. This coupling enables layer-controlled, Q1D spin transport with fully spin-polarized currents along orthogonal directions, as demonstrated in monolayer Mg$_2$Mo$_2$(PO$_5$)$_2$ and related candidates. The same RSLC mechanism further gives rise to a layer-dependent electric Hall effect, offering a magnetic-field-free route for Hall current generation. Our findings establish RSLC as a fundamental mechanism for spin-dependent physics in altermagnets and open the door to ridgetronic devices that harness dispersionless electronic states for next-generation spintronic functionalities.

\bibliography{ref}



\end{document}